# Deep Learning based Multi-modal Computing with Feature Disentanglement for MRI Image Synthesis

Yuchen Fei[1], Bo Zhan[1], Mei Hong[1], Xi Wu[2], Jiliu Zhou[1,2], Yan Wang[1, *]

[1] School of Computer Science, Sichuan University, China.
[2] School of Computer Science, Chengdu University of Information Technology, China.

*Corresponding author. E-mail: wangyanscu@hotmail.com

## Abstract

**Purpose:** Different Magnetic resonance imaging (MRI) modalities of the same anatomical structure are required to present different pathological information from the physical level for diagnostic needs. However, it is often difficult to obtain full-sequence MRI images of patients owing to limitations such as time consumption and high cost. The purpose of this work is to develop an algorithm for target MRI sequences prediction with high accuracy, and provide more information for clinical diagnosis.

**Methods:** We propose a deep learning based multi-modal computing model for MRI synthesis with feature disentanglement strategy. To take full advantage of the complementary information provided by different modalities, multi-modal MRI sequences are utilized as input. Notably, the proposed approach decomposes each input modality into modality-invariant space with shared information and modality-specific space with specific information, so that features are extracted separately to effectively process the input data. Subsequently, both of them are fused through the adaptive instance normalization (AdaIN) layer in the decoder. In addition, to address the lack of specific information of the target modality in the test phase, a local adaptive fusion (LAF) module is adopted to generate a modality-like pseudo-target with specific information similar to the ground truth.

**Results:** To evaluate the synthesis performance, we verify our method on the BRATS2015 dataset of 164 subjects. The experimental results demonstrate our approach significantly outperforms the benchmark method and other state-of-the-art medical image synthesis methods in both quantitative and qualitative measures. Compared with the pix2pixGANs method, the PSNR improves from 23.68 to 24.8. Moreover, the ablation studies have also verified the effectiveness of important components of the proposed method.

**Conclusion:** The proposed method could be effective in prediction of target MRI sequences, and useful for clinical diagnosis and treatment.

**Key words:** Magnetic resonance imaging (MRI), Image synthesis, Generative adversarial networks (GANs), Deep learning.

# Contents

1# I. Introduction

Medical imaging technology refers to a noninvasive way to obtain contrast images of the internal tissue structures in the human body. Imaging techniques such as Computed Tomography (CT), Positron Emission Tomography (PET), and Magnetic Resonance Imaging (MRI) have been widely applied in clinical practice with crucial information for disease diagnosis and treatment [1]. Particularly, MRI has been adopted as the primary imaging modality in neuroanatomy [2] for its superiority in multimodal imaging and clear display of soft tissue structures in the human body. By using different pulse sequences and scanning parameters, multiple MRI modalities corresponding to the same anatomical structure can be obtained. Pathological features exhibited by different modalities are specific. For example, the T1-weighted (T1) modality obtained by pulse sequences magnetization preparative gradient echo (MPRAGE) and spoiled gradient recalled echo (SPGR) is used to observe anatomical structures for its clear visualization of gray matter and white matter tissues. The T1-with-contrast-enhanced (T1c) modality is adequate for observation of vascular structures and the blood-brain barrier. For observation of pathological tissue, the T2-weighted (T2) modality is generated by applying a single long Time of Repetition (TR) and long Time of Echo (TE) sequence. The fluid attenuation inversion recovery (FLAIR) modality is adopted for suppressing cerebrospinal fluid signals, thereby producing lesion enhancement. Multi-modal MRI images present complementary pathological information, which is of great help to clinical diagnosis [3].

To improve diagnostic accuracy and treatment efficiency, multi-modal MRI images are usually preferred in clinical practice. However, it may not be possible to obtain full-sequence MRI images of patients due to time consumption and high cost in clinical practice [4]. Concretely, in terms of the imaging of brain, a typical brain T2/FLAIR scan would take 5 minutes or more, while for T1 the time can also take a couple of minutes. On one hand, the high acquisition time may cause difficulty for children, obese individuals and those with respiratory compromise since they are required to lie still and hold their breath [36]. On the other hand, the image quality can be reduced if the acquisition time is shortened. Furthermore, the noise and metal artifacts may prohibit the acquisition of intact sequences. Loss of some MRI modalities or sequences may cause the increased inaccuracy diagnosis. Accordingly, synthesizing missing or corrupted modalities from acquired MRI modalities has potential value for improving the efficiency and accuracy of diagnosis and treatment.

In the past decades, a range of single-modality synthesis methods have been presented to synthesize the required target MRI modality and can be divided into two broad categories: the atlas-based matching method [5-9] and intensity transformation method [10-11]. In the atlas-based matching field, Hofmann et al. [5] developed an approach utilizing local pattern recognition and atlas registration for attenuation correction. Burgos et al. [9] considered a strategy that utilizes multiple atlas alignment and intensity fusion for synthesizing CT images from corresponding MRI images. In regard to the intensity transformation direction, Jog et al. [10] proposed a data-driven approach to image synthesis, which utilizes a trained bagged ensemble of regression trees to employ the synthesis transformation on image patches. Huynh et al. [11] proposed a structured random forest model to synthesize CT from MRI in a nonlinear manner.

As a matter of fact, single-modality input has shown its limitations in offering feature and pathological information, while the unique information from different MRI modalities is complementary to each other. Therefore, multi-modal MRI image synthesis has aroused increasing attention for the improvement of the synthetic image quality and diagnostic utility [12-15]. In this regard, Jog et al. [15] proposed a method based on traditional random forest regression (REPLICA), which utilized nonlinear regression, multi-scale information and contextual description to enhance image synthesis performance. Nevertheless, the approach in the paper has failed to consider the loss of high-frequency spatial information due to processing image patches of different spatial scales separately and averaging the predicted values.

During the past decade, deep learning methods represented by Convolutional Neural Network (CNN) have drawn a tremendous amount of attention in medical image processing for their excellence in learning features automatically [16-25]. Xiang et al. [18] proposed a deeply embedded CNN network that embeds temporary CT prediction images during the forward extraction of feature maps, which showed great performance in improving the quality of the synthetic image. Chartsias et al. [20] conducted a modality-invariant latent representation (MILR) model based on the encoder-decoder structure which extracted potential representations of various MRI modalities for fusion, and the target modality was synthesized by the decoder. Unfortunately, the original CNN based synthesis methods have a drawback that the context information is ignored, limiting the size of the receptive field. Conversely, the Full Convolutional Neural Network (FCN) demonstrates the effectiveness in preserving the image structural information. For example, Wei et al. [26] designed a 3D FCN model to synthesize FLAIR from other MRI modalities, with great assistance for the treatment of multiple sclerosis.

Recently, the Generative Adversarial Networks (GANs) have gained widespread attention for its state-of-the-art performance in capturing high spatial frequency information by applying constraints of adversarial loss since its introduction [27]. Generally, GANs are composed of two networks, namely, a generator to generate realistic data that approximates the real data distribution and a discriminator to distinguish the generated data from the ground truth. On account of the excellent feature extraction performance of GANs, multiple variants based on GANs have successfully been applied for medical image synthesis [21-23, 28-31]. In the single-modality synthesis field, Nie et al. [28] proposed a context-aware GANs for synthesizing CT images from MRI, which showed superior performance than both the atlas-based and random forest-based methods. Along the multi-modal synthesis direction, Olut et al. [22] proposed a GANs-based framework to synthesize Magnetic Resonance Angiograph (MRA) from T1 and T2 modalities by utilizing Huber loss to optimize the synthetic images. In [31], a dual-discriminator adversarial learning model with attention mechanism for T2 modality synthesis from T1 was proposed, and the subtle image details could be faithfully preserved.

Currently, most of the image synthesis methods including both single- and multi-modal ones process and extract features from the entire input image directly. As indicated in [32], however, images can be further decomposed into modality-invariant space with shared information and modality-specific space with specific information. Specific to the MRI image, we refer to the contour and architecture information shared by different modalities as the shared information, and the image texture, color, and contrast information associated with a particular modality as the specific information. Nevertheless, shared and specific information belong to different spatial dimensions, using an encoder of the same depth to extract both could possibly bring the problems of either absence of valid high-dimensional shared information or damage to low-dimensional specific information, leading to undesirable synthetic results. Theoretically, more representative features could be acquired if the mixed shared information and specific information are learned separately [32].

In this paper, inspired by the success of GANs and to make the most of complementary information from diverse MRI modalities, we propose a novel GANs based multi-modal computing model for MRI image synthesis with feature disentanglement strategy and the Cat-Conv fusion mechanism. The novelties and contributions of the paper can be summarized as follows.

- For processing complementary information from different modalities effectively, our model provides a feature disentanglement strategy that decomposes the images of different modalities into a shared space with modality-invariant information (i.e., the shared information) and a specific space with modality specificity information (i.e., the specific information). It enables the distribution of different features to be learned pertinently and contributes to the improvement of the synthetic image quality.

- Different from other multimodal synthesis methods, we adopt the concatenating first and then convolution (Cat-Conv) scheme to fuse the shared information solely. With this strategy, the ability for shared information to express its representative features can be further enhanced, which is of assistance in the decoding phase.

- To address the issue that the target modality is invisible during the testing phase, we propose a local adaptive fusion (LAF) module to simulate a pseudo-target modality close to the ground truth for the convenience of capturing shared information in the testing phase. Moreover, L1 loss is applied to make the pseudo-target modality more realistic so that more detailed texture features can be preserved.

- We also incorporate an intra-layer fusion method into the model, in which the Adaptive Instance Normalization (AdaIN) layer is introduced to combine the shared information and specific information in the decoding phase with the affine parameters replaced by the standard deviation and mean of specific information.

The rest of this paper is organized as the following sections: In Section II, we introduce the proposed MRI synthesis model architecture and methodology in detail. After that, we elaborate the setup of the experiment and conduct the comparative experiments to prove the superiority of our method in Section III. Ultimately, we discuss and summarize this paper in Section IV and Section V, respectively.

## II. Methodology

In this section, we present the proposed GANs model utilizing feature disentanglement and Cat-Conv fusion mechanism for effective synthesis of multi-modal MRI images. Figure 1 illustrates the structure of the proposed method. The framework is composed of two networks: the generator $G$ and the discriminator $D$. $G$ aims to generate a realistic target MRI modality that resembles the real image, adopting the encoder-decoder framework as the backbone. Meanwhile, $D$ learns to distinguish the synthetic image from the real one, utilizing the traditional CNN architecture to build the network.

Concretely, the encoders receive multiple MRI images as input and capture the abstract features by compressing the image resolution and enlarging the receptive field, while the decoder takes the features extracted by the encoders as input and restores the fused information with an up-sampling operation, finally generating the estimated MRI modality. Note that, in order to extract shared and specific information from different modalities, we consider two distinct encoders for effective feature extraction. One is the ShaRed Encoder (SRE), which is applied to extract high-dimensional shared information relevant to image semantics and overall structure. The other, SPecific Encoder (SPE), is employed to obtain low-dimensional specific information associated with the detailed texture of the image.

Since the shared information is highly similar in distribution, it is rational to fuse them to enhance the expression of the overall semantic features. Differently, the specific information must be extracted from the image that resembles the real target modality, but the target modality is not available during the testing phase. To deal with this problem, the local adaptive fusion (LAF) module is presented to simulate a pseudo-target modality which will be fed into the SPE to learn the specific information. Finally, the shared and specific information are fused in the decoding path, then the fused features are gradually restored to the synthetic target modality of the same size as the input modality by the up-sampling operation.

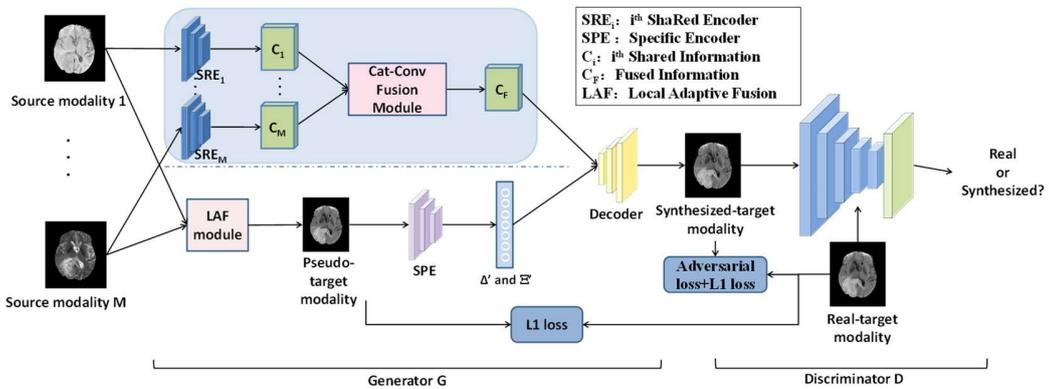

Figure 1: Framework of the proposed method.

The input of the discriminator $D$ is a set of images including the source MRI images and the corresponding real/synthetic target image. The multi-layer CNN is then applied to acquire the feature map with a global receptive field, followed by the sigmoid activation to determine whether the image is real or synthesized. The architectures of generator and discriminator are presented in the following Sections II-A and II-B, and the objective functions and implementation details of our model are discussed in Section II-C and Section II-D, respectively.

## II.A. Generator network

### II.A.1. Encoder Structure

***ShaRed Encoder (SRE):*** Since the generator network receives multiple MRI images with different contrasts as input, it is reasonable to set a corresponding encoder, i.e., SRE, to extract the shared information of each input modality, which is identical to the fact that the number of SRE equals to that of the input modalities as indicated in Figure 1. Notably, each SRE shares the same structure but learns different parameters.

In our model, each SRE consists of three down-sampling blocks and four residual blocks (ResBlock). Specifically, we design the down-sampling block as a "Convolutional Layer (Conv) - Instance Normalization Layer (IN) - ReLU Activation Layer (ReLU)" structure. The input image of size 256×256 is first convolved by a 7×7 kernel with a padding of 3 and a stride of 1, to increase the number of channels while keeping the scale constant. In the following two down-sampling blocks, the kernel size is set to 4×4 with stride 2 and padding 1 to halve the feature map size and increase the number of channels. Particularly, we adopt the instance normalization instead of batch normalization to maintain the independence of each modality, thus ensuring the stylization of image is not affected by the batch size. In the residual blocks, the "Padding - Conv - IN - ReLU" structure is employed to deepen the network and extract high-dimensional shared information, while avoiding the network optimization problems caused by the deep network hierarchy. Moreover, the size of the feature map and the number of channels remain unchanged in the ResBlock. According to the above architecture, the output of the SRE is a feature map size of 256×64×64.

***SPecific Encoder (SPE):*** There are huge differences exhibited in function and structure between SPE and SRE. In function respect, it is necessary for SPE to learn specific information from the image that resembles the real target modality. As for the structure, as the shallower network layers are capable of extracting low-dimensional specific information relevant to image style, such as the edges and image textures, we design the SPE as a network with fewer layers. It is effective with five modules structured as "Conv-ReLU" to extract the specific information of each modality. In addition, for better fusion processing, the IN layer is removed for its poor performance in retaining the original features and standard deviations of the specific information [33]. The pseudo-target image of size 256×256 is passed into the first convolutional layer with a 7×7 kernel with stride 1 and padding 3, which is followed by four modules whose kernel size is set to 4×4 with stride 2 and padding 1. Subsequently, each channel of the feature block containing specific information is compressed into a real number with a global average pooling operation. After that, the feature channel dimension is reduced to 8 with a 1×1 convolution kernel, and through the last three linear layers we can finally obtain the mean $\Xi'$ and the standard deviation $\Delta'$ of the specific information which are utilized for fusion with the shared information in the decoding phase.

### II.A.2. Local Adaptive Fusion Module

To address the issue that the target modality is not available and facilitate the extraction of specific information during testing phase, we design a local adaptive fusion (LAF) module before SPE to synthesize a pseudo-target modality for acquiring the specific information. The architecture of the LAF module is provided in Figure 2.

Taking the dual-modality input as an example, we argue that the contribution to the same image region varies from one modality to another, and the contribution of the same modality to different regions may also be different. To make better use of the information across modalities, the multi-modal input images are partitioned into several image blocks in the same way for feature extraction. As shown in Figure 2, the regions from different modalities with the same color indicate that they correspond to the same location of the original image. The 1×1 convolution with 2 channels is adopted for regions at the same location, with different convolution kernels for different regions. The image blocks after convolution will then be combined according to their previous relative positions, thereby generating the pseudo-target modality for subsequent specific encoding. Note that the 1×1 convolution provides an end-to-end multi-modal fusion mechanism, in other words, each modality is weighted by the convolutional parameters to reflect the contributions of different modalities to different regions. In addition, the end-to-end training also reduce the number of parameters to be learned compared with the multi-channel cascade approach. Finally, we consider the L1 loss as the objective function to optimize the network, so as to generate a more realistic target modality with sharper information.

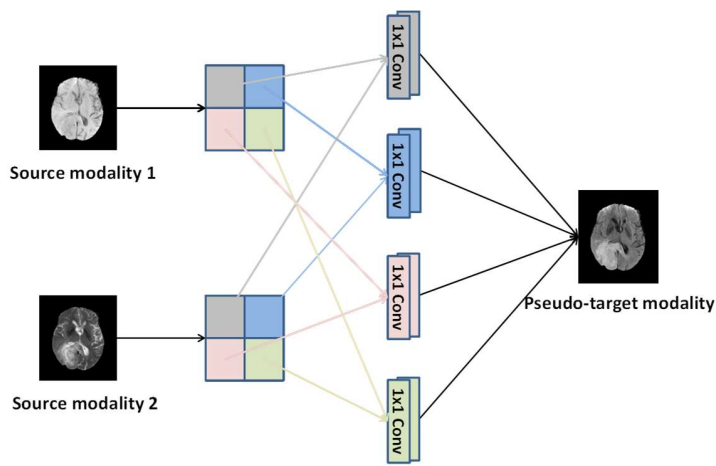

Figure 2: Structure of Local Adaptive Fusion (LAF) Module.

### II.A.3. Cat-Conv Fusion Module

In the shared information encoding path, the concatenating first and then convolution (Cat-Conv) approach has been proposed to fuse the shared information from multiple modalities. Concretely, the feature channels of the shared information extracted by SRE are concatenated first and then convolved through a convolutional block, compressed to effectively fuse the information from multiple modalities. The detailed architecture of this module is provided in Figure 3.

As observed, $\{C_1, \cdots C_M\}$ represent the shared features extracted by SRE from M source modalities with a size of 256×64×64 and are cascaded along the channel direction, thereby generating a feature with channel of 256×M. Specifically, we set a 3×3 kernel for convolution with a stride of 1 and a padding of 1 to process the shared information after cascading. All channels are weighted and summed in the convolutional layer, and finally the fused information $C_F$ with a channel of 256 is obtained. It is worth noting that the task of the Cat-Conv module is in charge of the enhancement of channel information solely, without reducing the resolution of abstract features in the high level, which is conducive to the further fusion with specific information.

### II.A.4. Decoder Structure

The decoder takes the shared information after Cat-Conv fusion as well as the means and standard deviations of the specific information as its input to synthesize the target modality. In this section, we propose an Adaptive Instance Normalization (AdaIN) layer to fuse the two types of information effectively in the decoder, where the specific information is embedded in the shared information by affine transformation.

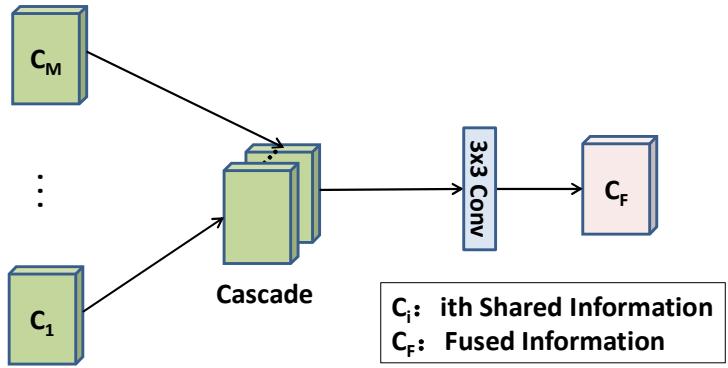

Figure 3: Structure of Cat-Conv Fusion Module.

Our decoder includes of four residual blocks, each of which adopts the structure similar to that of the residual blocks in SRE except that we replace the IN layer with the AdaIN layer for intra-layer fusion of shared and specific information. Moreover, the affine transformation parameters γ and β in the traditional IN layer are closely related to the style of the image and need to be learned through the training of the network, as shown below:

$$IN(x) = \gamma \left( \frac{x - \mu(x)}{\sigma(x)} \right) + \beta, \qquad (1)$$

where γ denotes the scale parameter and β denotes the offset parameter, $x$ is the input sample, $\mu(x)$ and $\sigma(x)$ represent the mean value function and standard deviation function, respectively. Since the style of the target modality has already been determined in our proposed model, the constant parameters $\Delta'$ and $\Xi'$ extracted from the specific information are considered to replace the affine parameters in the IN layer, and the computational burden can be further reduced to some extent. Thus, the expression of the AdaIN layer is described as:

$$\text{AdaIN}(x) = \Delta' \left( \frac{x - \mu(x)}{\sigma(x)} \right) + \Xi', \qquad (2)$$

where $x$ denotes the fused feature after the Cat-Conv fusion, $\Delta'$ denotes the standard deviation of the specific information output from SPE, and $\Xi'$ is the corresponding mean value. In the manner of fixing parameters of the transformation formula, the shared and specific information can be fused effectively without hard coding at the AdaIN layer. The feature after fusion through the four residual blocks is then up-sampled with three transposed convolution (ConvT) blocks to restore the high-dimensional feature map to an image with the same size as the source modality, finally the synthesized target modality image is acquired.

## II.B. Discriminator network

The discriminator adopts a network structure similar to the structure of binary classification CNN. The discriminator network is comprised of five convolution blocks, and receives a group of images including the source input MRI images and the corresponding real/synthetic target image as input. After the first layer of convolution, the number of channels of feature maps increases to 64 and is doubled in each subsequent convolution. The metric output by the last sigmoid activation function is employed to determine whether the image is real or synthesized, the closer it is to 1, the more likely the input target modality is true.

## II.C. Objective Functions

According to the above discussion, the objective function of the proposed approach consists of three loss functions in total, which can be summarized as follows: 1) the L1 loss between the synthetic target modality and the real target modality, which is employed to enhance the quality of the synthetic target modality; 2) the L1 loss between the pseudo-target modality and the real target modality, which enables the pseudo-target modality to retain more specific information; 3) the adversarial loss of the GANs, which aims to make the synthetic image more realistic to fool the discriminator. Illustrated by the example of the synthesis with two modalities as input, $X_1, X_2$ represent the images of the two input modalities, $Y$ is denoted as the image of the real target modality, $G$ and $D$ are the generator network and discriminator network, respectively, and $F$ denotes the local adaptive network. Consequently, $\bar{Y} = G(X_1, X_2)$ is the synthesized target modality, and the pseudo-target modality can be represented by $\bar{\bar{Y}} = F(X_1, X_2)$. The overall loss function can be expressed as:

$$L_{total} = arg\min_{G}\max_{D} L_{cGAN}(G, D) + L_1(G, F). \tag{3}$$

In Eq. (3), $L_{cGAN}(G, D)$ is formulated by:

$$L_{cGAN}(G, D) = \mathbb{E}_Y[log(D(Y))] + \mathbb{E}_{X_1, X_2}\left[1 - log\left(D(G(X_1, X_2))\right)\right]. \tag{4}$$

$L_1(G, F)$ is calculated as follows:

$$\begin{aligned} L_1(G, F) &= \lambda_1 \mathbb{E}_{X_1, X_2, Y}[\|Y - \bar{Y}\|_1] \\ &+ \lambda_2 \mathbb{E}_{X_1, X_2, Y}\left[\left\|Y - \bar{\bar{Y}}\right\|_1\right] \\ &= \lambda_1 \mathbb{E}_{X_1, X_2, Y}[\|Y - G(X_1, X_2)\|_1] \\ &+ \lambda_2 \mathbb{E}_{X_1, X_2, Y}[\|Y - F(X_1, X_2)\|_1]. \end{aligned} \tag{5}$$

With regard to Eq. (4), as discussed above, $G$ is expected to generate a high-quality image that fools $D$ into giving a prediction as close to 1 as possible, which is consistent with the maximization of $D(G(X_1, X_2))$. As for the discriminator $D$, it tries to maximize $D(Y)$ while minimize $D(G(X_1, X_2))$, so that the real modality is distinguished from the generated image. In order to make the synthesized image more realistic, the pixel-wise differences between the real target image and the synthesized target image as well as the pseudo-target image are calculated in Eq. (5) and minimized in Eq. (3). The parameters $\lambda_1$ and $\lambda_2$ are employed to balance the two items of L1 loss, both of which are empirically set to 0.1 in the experiment.

## II.D. Implementation Details

Experiments are conducted utilizing the PyTorch framework with the NVIDIA GeForce GTX 1080Ti with 11GB memory. Specifically, we apply 200 epochs to train the proposed model utilizing the Adam optimizer with a batch size 3. In the first 100 epochs, the learning rates for both generator and discriminator networks are fixed as 0.0002, then linearly decay to 0 in the next 100 epochs. The instance normalization is applied in the SRE and the adaptive instance normalization is adopted in the decoder, and batch normalization is employed for discriminating phases. Furthermore, the Leaky ReLU layer with the slope of 0.2 is introduced for activation in each layer. In terms of the original GANs [27], we train the generator and discriminatory networks in an alternating manner. Generally, $G$ is first fixed to train $D$, and then $D$ is fixed to train $G$ for each epoch.

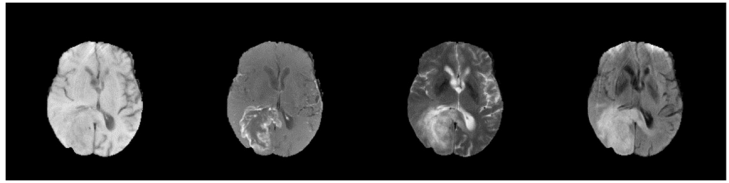

Figure 4: From left to right: T1, T2, T1c and FLAIR modalities.

## III. Experiments and Results

### III.A. Dataset

We evaluate the performance of our proposed method on the official dataset from the Multimodal Brain Tumor Segmentation (BRATS) Challenge [34] in 2015, which contains co-registered MRI images collected from 124 subjects with high-grade glioma and 40 subjects with low-grade glioma, grouped as: T1, T1c, T2 and FLAIR. Each modality volume is of the same size 155×240×240 (voxels). Considering the small number of training samples, we perform slice preprocessing along the image axial direction and each 3D scan is split into 2D axial-plane slices with a size of 240×240. The slices without pixels are firstly excluded from the entire slice sequences, resulting in approximately 120 slices for each subject. Then, 80 slices in the middle of the sequence of each subject are selected to the training or testing cohort. Each slice is further resized to 256×256 to accommodate to the input of our model. In the experiments, 126 subjects with 10080 slices are randomly selected for training the model, while the remaining 38 subjects with 3040 slices for testing. For each epoch, it takes about 10 minutes to train, and the testing phase takes about 2.5 minutes. Some slices from the dataset are displayed in Figure 4.

### III.B. Experimental Setting

To study the effectiveness of the proposed multi-modal synthesis method, we conduct a set of experiments in three aspects. First, we evaluate the performance of the key component, which includes verifying the selection of chunk size and the validity of the LAF module. The results are presented in both qualitative and quantitative measures in Section III-D. Next, we investigate the effectiveness of multi-modal input in Section III-E, by comparing the synthesized images generated by the model with single-modality input, double-modality input and triple-modality input. Finally, we evaluate the overall performance of our model in comparison with the benchmark model and other state-of-the-art methods, which is shown in Section III-F.

## III.C.  Metrics of Performance

For quantitative evaluation of the performance and effectiveness of our model, we employ three standard metrics in the experiments to measure the performance of our model and other methods in comparison: peak signal-to-noise ratio (PSNR), structural similarity index (SSIM) and normalized root-mean-square error (NRMSE).

PSNR is defined as:

$$PSNR = 10 log_{10} \left( \frac{A \times MAX_Y^2}{\|Y - Y'\|_2^2} \right) \quad (6)$$

where $Y$ illustrates the real target image and $Y'$ refers to the synthesized target image, $A$ represents the total number of pixels, and $MAX_Y$ denotes the maximum intensity value of the image. PSNR is applied to measure the quality of the synthesized image, with values ranging from 20 to 40, and higher PSNR demonstrates better synthesis performance.

SSIM is computed as:

$$SSIM(Y, Y') = \frac{(2\mu_Y \mu_{Y'} + C_1)(2\sigma_{YY'} + C_2)}{(\mu_Y^2 + \mu_{Y'}^2 + C_1)(\sigma_Y^2 + \sigma_{Y'}^2 + C_2)} \quad (7)$$

where $\mu_Y$ refers to the means of $Y$ and $\mu_{Y'}$ denotes the means of $Y'$, $\sigma_Y$ and $\sigma_{Y'}$ represent the variances of $Y$ and $Y'$, respectively, and $\sigma_{YY'}$ is the covariance of $Y$ and $Y'$. The positive constants $C_1$ and $C_2$ are introduced to avoid problems caused by the null denominator. SSIM is employed to evaluate the structural similarity between two images. Generally, the closer the value of SSIM is to 1, the more similar the structures of the real and synthesized images will be.

NRMSE is calculated as:

$$NRMSE(\hat{Y}, Y) = \sqrt{\frac{\sum_{x=1}^{N}(\hat{Y}(x) - Y(x))^2}{\sum_{x=1}^{N} Y(x)^2}} \quad (8)$$

NRMSE is exploited to measure the difference between the real images and synthesized images. Notably, different from PSNR and SSIM, a synthesized image with lower NRMSE indicates higher synthesis quality.

According to the above equations, it can be summarized that synthetic images with higher PSNR and SSIM values, lower NRSME values show superior image quality and demonstrate better performance of the method.

## III.D.  Validation of LAF Module

In Section II-C-3, the local adaptive fusion (LAF) module is proposed to fuse multi-modal images into a pseudo-target modality analogous to the real target image in the testing phase, in which each input image is required to be divided into blocks of the same size in a same way and convolved with different convolution kernels.

To evaluate the effect of block size on the fusion performance, we design a set of comparison experiments for different block sizes with the synthesis of FLAIR modality from T1 and T2 as an example. T1 and T2 modality images are divided into 2×2, 4×4, 8×8, and 16×16 blocks with corresponding block sizes of 128×128, 64×64, 32×32, and 16×16, while the block of size 256×256 without chunking is adopted as the benchmark for comparison. It is worth noting that the group without chunking is in essence fused directly by cascading and then convolution, which is equivalent to the group that are not processed by the LAF module. In this manner, this experiment is employed to not only validate the selection of the optimal chunking size but also evaluate the contribution of the LAF module to the performance of the proposed model. The qualitative and quantitative experimental results are presented in Figure 5 and 6 respectively.

From the regions of real and synthetic FLAIR modalities indicated by the red boxes, it can be observed that the chunking size in LAF module makes an actual effect on the quality of the synthesized image. Among the partitioned groups, the FLAIR modality synthesized by the group with a block size of 128×128 is most analogous to the real image and preserves more details compared with other synthetic results. The quantitative comparison results of groups with various block sizes in terms of PSNR, SSIM and NRMSE is illustrated by Figure 6. It can be found that the synthesis ability of our model shows a tendency to first rise and then decline with the decreasing size of chunking. Specifically, the group with block size of 128×128 offers the optimal or near-optimal performance, which is identical to the visualization results in Figure 5. We reason that this trend may be related to the loss of neighborhood information between blocks after image partitioning. When the size of block is kept as 256×256, a same convolutional kernel has to be applied in different image regions, which shows limitations in extracting the effective features though the information of the whole image is well retained. While the slight loss of neighborhood information of the image blocks has little impact on feature extraction when the block size is reduced to 128×128. Moreover, extracting features from different regions by distinct convolutional kernels is conductive to the learning of specific information in a targeted way, which contributes to the expression of valid information in the image. With the decrease of the block size, the correlations between blocks and the neighborhood information embedded in the image blocks are significantly reduced, which gives rise to the very limited information in learned features and reduced quality of the synthesized image.

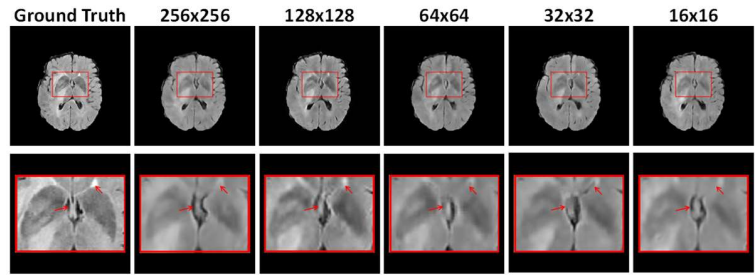

Figure 5: Visualization comparison of synthesized images generated by different block sizes.

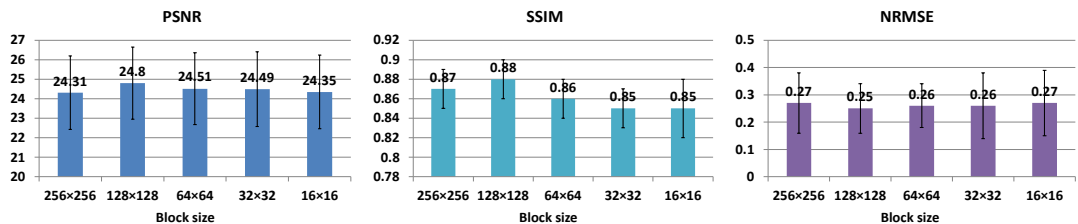

Figure 6: Quantitative comparison of synthesized images generated by different block sizes. (T1+T2→FLAIR).

In Summary, it is verified that the highest synthetic quality is obtained when the block size is 128×128. Moreover, both qualitative and quantitative results demonstrate the benefits of employing the proposed LAF module in enhancing the synthesis performance. Similarly, it also produces the optimal result in terms of NRMSE. As for the SSIM metric, the T1+T2→FLAIR group yields the highest score, however, there is no significant difference among the three experimental groups. According to both the qualitative and quantitative experimental results, we can infer that the image synthesis performance of our model in image clarity and detail can be enhanced as the number of input modalities increases.

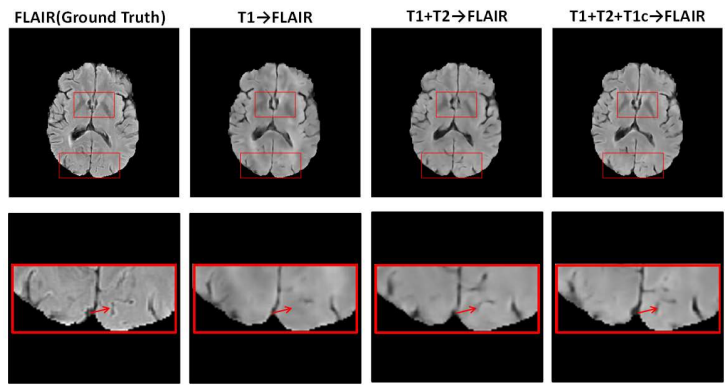

Figure 7: Visualization comparison of synthesized images by the proposed model with different multi-modality inputs.

Table 1: Qualitative comparison of synthesized images by the proposed model with different multi-modality inputs.

| FLAIR synthesis | PSNR | SSIM | NRMSE |
| --- | --- | --- | --- |
| T1→FLAIR | 23.7±2.16 | 0.86±0.02 | 0.30±0.11 |
| T1+T2→FLAIR | 24.8±1.85 | **0.88±0.02** | 0.25±0.09 |
| T1+T2+T1c→FLAIR | **24.93±1.96** | 0.87±0.02 | **0.24±0.11** |

## III.E. Effectiveness of Multi-modal Input

Based on the theory that increasing the input modalities can enrich the features learned by the model and thus improve the quality of the synthesized images, we propose a multi-modal image synthesis model. In this section, to evaluate the effectiveness of the multi-modal input, we conduct the comparison experiments that respectively take T1, T1+T2, and T1+T2+T1c as input of our model to synthesize the FLAIR modality. In addition, for the purpose of ensuring the single variable principle of experiments, only one more input modality will be incorporated into the current model with respect to the previous one. The qualitative and quantitative results are presented in Figure 7 and Table 1, respectively.

From Figure 7, it can be observed that the sharpness of textures in the regions marked by the red boxes is enhanced and the synthetic image more closely resembles the real FLAIR modality image as the number of input modalities increases. As shown in Table 1, the triple-input model T1+T2+T1c→FLAIR achieves the highest PSNR values, with 1.23dB higher than that of the single-modality input group T1→FLAIR and 0.13dB higher than that of the dual-modality group T1+T2→FLAIR. Similarly, it also produces the optimal result in terms of NRMSE. As for the SSIM metric, the T1+T2→FLAIR group yields the highest score, however, there is no significant difference among the three experimental groups. According to both the qualitative and quantitative experimental results, we can infer that the image synthesis performance of our model in image clarity and detail can be enhanced as the number of input modalities increases.

## III.F. Comparison with Existing State-of-the-art Methods

In addition to verifying the validity of the sub-modules of the proposed method, we also evaluate the overall performance of the proposed model by comparing it with other existing state-of-the-art methods, including: 1) pix2pix GANs utilizing U-net [35] as the generator (Baseline); 2) Modality-Invariant Latent Representation (MILR) [20]; 3) Multi-contrast steerable Generative Adversarial Networks (sGANs) [22].

It is worth mentioning that there are some implicit confirmatory experiments in the comparisons. Concretely, the difference between pix2pix GANs and our method is the structure of generator. For pix2pix GANs, U-net is employed as backbone which stacks multiple modalities directly and treats them as a multi-channel image to handle. In contrast, our method applies various encoders to extract distinct shared features and specific features from multiple input modalities separately, which is regarded as feature disentanglement strategy. Meanwhile, although MILR and our method both allow for the possibility of feature disentanglement, MILR mainly concentrates on the extraction and transformation of shared information without including specific information while our method considers both of them. Hence the comparison between multi-channel pix2pixGAN and our method can demonstrate the effectiveness of our proposed feature disentanglement. Similarly, we can futher study the validity of the introduction of specific information by comparing our method with MILR.

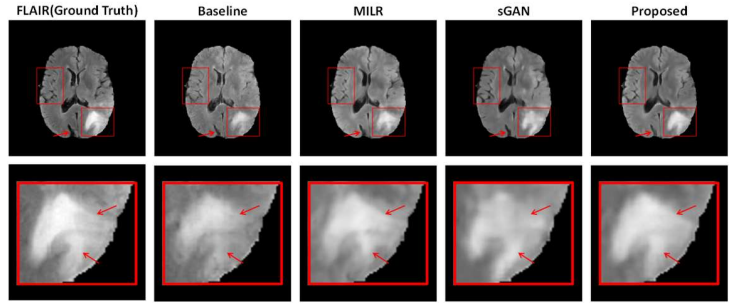

Figure 8: Visual comparisons of synthesized images by Baseline, MILR, sGANs and the Proposed method.

Table 2: Qualitative comparisons of synthesized images by Baseline, MILR, sGANs and the Proposed method. * means our method is significantly better than the compared method with p<0.05 via paired t-test.

| T1+T2→FLAIR | PSNR | SSIM | NRMSE |
|---|---|---|---|
| pix2pix GANs | 23.68±2.03* | 0.86±0.03* | 0.3±0.12* |
| MILR | 24.49±1.92* | 0.87±0.02* | 0.27±0.10* |
| sGANs | 24.52±1.97* | 0.87±0.02* | 0.26±0.11 |
| **Proposed** | **24.8±1.85** | **0.88±0.02** | **0.25±0.09** |

Taking the synthesis of FLAIR from T1+T2 as an example, the method presented in this paper is denoted as Proposed, and the visual and quantitative comparison results are given in Figure 8 and Table 2 separately.

The major differences of the images generated by the four synthetic methods have been marked with red boxes and arrows in Figure 8. As observed, the Baseline obtains the worst synthetic results, with the most blurred structure in the white lesion region and least texture information. The synthetic results of sGANs show more sharper textures than that of MILR, while our proposed method can generate the synthesized image with the appearance mostly resembles the ground truth. For quantitative comparison in Table 2, the results have demonstrated that our proposed method outperforms other existing methods with the highest PSNR and the lowest NRMSE. Although the SSIM has not been significantly improved, the comparison experiments still demonstrate the effectiveness of the proposed feature disentanglement strategy compared with the benchmark method and the superiority of synthetic performance over other state-of-the-art methods. Moreover, we have also conducted t-test to study if our method statistically significantly outperforms other methods. Through paired t-tests, the p-values of all the metrics are consistently smaller than 0.05, demonstrating the improved synthesis performance generated by our method is statistically significant.

## IV. Discussion

In view of the limitations of time constraints and high cost of MRI images in clinical practice,

we propose a novel GANs based method to generate the missing or damaged modality from multi-modal MRI images. Inspired by the remarkable success of image style transformation and to overcome the deficiencies of existing multi-modal synthesis methods, we consider a feature disentanglement strategy to decompose the image into the shared space with modality-invariant information and the specific space with modality-specific information. For specific information, the LAF module is proposed to simulate a pseudo-target modality for the subsequent extraction of the mean and standard deviation by the SPE, while the shared information extracted by the SRE is integrated by the Cat-Conv module. Next, the decoder employs the AdaIN layer to restore the integrated information with an up-sampling operation, and the target MRI modality is finally obtained. Last but not least, the discriminator determines whether the input target modality is real or not and provides a reference for improving the quality of the synthesized image. In addition, L1 loss is considered as the objective function to further optimize this model.

Extensive experiments are conducted on the BRATS2015 dataset to measure the performance and efficiency of our model, with PSNR, SSIM, NRMSE as metrics. As discussed in Section III-D, we verify that the best synthesis effect is obtained when the block size is 128×128, and both qualitative and quantitative results have demonstrated the benefits of employing the proposed LAF module in enhancing the synthesis performance. For the purpose of studying the effectiveness of the multi-modal input, we separately take T1, T1+T2, T1+T2+T1c as input to synthesize the FLAIR modality. From Section III-E, it can be found that there is significant improvement in the quality of the synthesized image as the number of input modality increases. At the same time, to investigate the overall performance of our model, we compare the proposed method with the benchmark method based on U-net and other state-of-the-art methods MILR and sGANs. Obviously, our method generates the target modality most similar to the real image and achieves best performance with the highest PSNR, SSIM and the lowest NRMSE. Moreover, we also verify the effectiveness of the feature disentanglement strategy and superiority in extracting specific information through comparison with pix2pixGAN and MILR.

Although our method outperforms other state-of-the-art methods in both visual and quantitative evaluation, there are still several limitations of our model. One major limitation is that the improvement is still slight especially for SSIM which indicates the similarity of the structure of two images. What's more, since the error punishment for the edge information is absent in the objective function, the tumor boundary is still less sharp. Consequently, we consider extending our method and introducing some strategies that are expedient in preserving sharper textures in our future work, for example, employing Huber loss or image gradient loss apart from L1 loss, and further optimizing the LAF module and specific encoder to obtain the pseudo-target image with more edge details and specific information of high accuracy. For shared information fusion, the Cat-Conv module can be further designed with a channel attention mechanism to enhance the feature expression of different channels. In addition, the current study is restricted to 2D images, which inevitably causes the loss of context information between slices, thereby limiting the extraction of features. Also, there is only a limited number of MRI images available to train and evaluate the proposed method. In the future, we will involve more subjects into the study and conduct a variety of studies with 2.5D or 3D input to preserve more detailed information for synthesizing images of high quality. Furthermore, the synthesized MRI image can be applied in the medical image analysis tasks such as segmentation and classification. In light of this, we will consider combining our method with downstream tasks to boost the synthesis performance.

# V. Conclusion

In this paper, we propose a new deep learning method based on GANs to synthesize the target MRI modality from multi-modal input. The feature disentanglement strategy is introduced to separately process the specific and shared information from different modalities, and the Cat-Conv and LAF modules are applied for the fusion of shared information and synthesis of pseudo-target modality, respectively. Moreover, the introduced AdaIN layer can integrate the specific and shared information effectively, which is an improvement on existing image synthesis methods. Extensive experiments have been conducted on the BRATS dataset, and our proposed method is demonstrated outperform state-of-the-art methods by both qualitative and quantitative results. In our future work, we will try to exploit more detail retention techniques to further extend our method for superior performance.

## Acknowledgment

This work is supported by National Natural Science Foundation of China (NSFC 62071314) and Sichuan Science and Technology Program (2021YFG0326, 2020YFG0079).